\documentclass[]{aa}
\input{epsf}
\input{psfig}
\def\kms{~km~s$^{-1}$}

\begin{document}
\thesaurus{11.03.1; 11.03.4; 11.04.1; 11.11.1; 12.03.3}
\title{ Abell 521: Dynamical analysis of a young cluster
\thanks{Based on observations made at the Canada France Hawaii Telescope (CFHT)
and at the European Southern Observatory. CFHT is operated by the National
Research Council of Canada, the Centre National de la Recherche Scientifique
of France , and the University of Hawaii.  }} %
\author{
S.Maurogordato\inst{1}, D.Proust\inst{2},  T.C. Beers \inst{3}, M. Arnaud
\inst{ 4}, R. Pell\'o \inst{5}, A.Cappi\inst{6}, E.Slezak \inst{1}, J.R.
Kriessler \inst{ 7}} %
\institute{
$^1$ Observatoire de Nice, B4229, Le Mont-Gros, 06304 Nice Cedex 4,
E-Mail: maurogor@obs-nice.fr \\
$^2$ DAEC, CNRS, Observatoire de Paris-Meudon,
5 place J. Janssen, 92195 Meudon, France,
E-Mail: proust@obspm.fr \\
$^3$ Department of Physics \& Astronomy, Michigan State University,\\
E. Lansing, MI  48824, USA, Email: beers@pa.msu.edu \\
$^4$ CEA/DSM/DAPNIA/Service d'Astrophysique, CEA/Saclay, F-91191
   Gif sur Yvette Cedex, France
E-Mail arnaud@hep.saclay.cea.fr \\
$^5$ Observatoire Midi-Pyr\'en\'ees, Avenue E.Belin, Toulouse
E-Mail: roser@astro.obs-mip.fr\\
$^6$ Osservatorio Astronomico di Bologna,
via Zamboni 33, I--40126 Bologna, Italy,
E-Mail: cappi@astbo3.bo.astro.it \\
$^7$  Department of
Astronomy, University of Minnesota, 116 Church St. S.E.,
Minneapolis, MN 55455, USA,
Email: jeffk@isis.spa.umn.edu
}
\offprints{Sophie Maurogordato}
\date{Submitted~ 1999 January 21 }
\maketitle
\markboth{Maurogordato et al.}{Abell 521: Dynamical analysis of a young cluster}
\begin{abstract}

We present the results of a dynamical analysis of the rich X-Ray luminous
galaxy cluster Abell 521, and discuss the nature of the arc-like structure
first noted by Maurogordato et al. (1996). Our study is based on radial
velocities for 41 cluster members, measured from spectra obtained at the
European Southern Observatory and the Canada-France-Hawaii Telescope.  Based on
statistical analyses performed with the ROSTAT package, we find that Abell 521
is an intermediate-redshift cluster ($C_{BI}= 74132 _{-250} ^{+202}$ km/s) with
a rather high apparent value of the velocity dispersion $S_{BI}= 1386 _{-139}
^{+206}$ km/s.
 
There are many indications that this cluster is presently undergoing strong
dynamical evolution: a) the high value of the velocity dispersion, which cannot
be explained by trivial projection effects, b) signific ant clumping in the
two-dimensional projected positions of the galaxies in the cluster, quantified
by a mixture-model three-group partition significant at 99 $\%$ level, c) the
extreme value of the velocity dispersion ($\sigma \sim 2000$ km/s) in a central
high density NE/SW  structure,  d) a strong increase of the velocity dispersion
as determined from the reddest and bluest galaxies, suggesting that cluster
spirals are not yet virialized; e) the presence of multiple nuclei in the core
of the brightest cluster galaxy , one of which has a statistically significant
velocity offset from the rest of the cluster members, and f) an apparently
different stellar population for the various knots of the arc candidate which
changes along the structure.
 
The two brightest knots of the giant arc candidate are shown to be at the
velocity of the cluster.  This makes the gravitational lensing interpretation
for the bright curved structure very improbable, although gravitational lensing
might  still be present in this cluster, as suggested by the colors of two
fainter arclet-like structures.

\end{abstract}
 
\section{Introduction}
 
Clusters of galaxies are complex, evolving systems which present numerous
observational and theoretical challenges, yet the scientific payoff of a
detailed understanding of these structures is also great.  The determination of
the total gravitationally-bound mass, the relative distribution of visible
(galaxies and hot gas) and dark matter, and the dynamics of galaxies in
clusters all provide essential information for testing models of galaxy
formation and evolution.
 
There exist numerous uncertainties in deriving the mass distribution within
clusters from optical observations alone (Merritt and Gebhardt 1995).  X--ray
observations of the hot gas in clusters permit significant progress in our
understanding of cluster dark matter distributions.  The two approaches are
complementary, but in both instance strong assumptions are required for
recovering the mass distribution.  In particular, the optical approach (through
the virial analysis) requires one to adopt hypotheses concerning the orbital
distribution of member galaxies, while the X--ray approach assumes that the hot
gas is in hydrostatic equilibrium within (presumably) a single cluster
potential well.  A comparison of both methods, when possible, gives the
opportunity to test the underlying hypotheses and to better constrain the model
parameters (Henry et al. 1993).  With this aim in mind, we have initiated a
combined X--ray and optical observational program, including both imaging and
multi--object spectroscopy at ESO and the CFHT, on a selected sample of
clusters at intermediate redshifts.  Our sample was defined in order to cover a
variety of clusters, which could be taken as representative of a range of
dynamical states.
 
Abell 521 is a distant (Abell distance class 6), relatively rich (Abell
richness class 1), southern cluster, morphologically classified as Bautz-Morgan
Type III (Abell 1958; Abell, Corwin, and Olowin 1989).   This cluster was shown
by the HEAO-1 survey to be a bright X--ray source (Kowalski et al. 1984).  In
images taken to select spectroscopic targets in the fields of this cluster, we
detected several arc candidates (Maurogordato et al.  1996).  This was not
surprising, as it is suggested by theoretical studies of the gravitational
lensing phenomenon that luminous arcs should be frequently found in the centers
of distant X--ray luminous clusters (Le F\`evre et al.  1995).  In our case,
the detection of luminous arcs, {\it if real}, is important because they
provide an independent means to measure the {\em total} mass of the cluster,
without requiring explicit models of the mass distribution (Fort and Mellier
1994). All this motivated us to focus on this cluster for deeper analysis,
mainly through multi-object spectroscopy, to probe its dynamical state,
long-slit spectroscopy to test the reality of the arc and arclet candidates,
and multi-band photometry of the central region, with X--Ray imaging and
spectroscopy being conducted in parallel.  A comparison of the X--Ray,
optical images and temperature measurement of the main cluster will be presented in Arnaud et al. (1999); a
photometric redshift analysis of the arc candidates from multicolor photometry
will appear in Pell\'o et al. (1999).
 
Section 2 of this paper presents the spectroscopic data we have obtained to
date in our observational campaign on this cluster.  Section 3 is an analysis
of the velocity distribution.  In section 4 we discuss the nature of the
brightest cluster galaxy, and the reality of the gravitational arc candidates
which have been suggested previously. In the following, unless explicitly specified, we have used $H_0 = 50 \rm h_{50} \rm km s^{-1} Mpc^{-1}$, and 
$q_0=0.5$.
 
\section{Observations and data reduction}
 
Multi-object spectroscopic observations of Abell 521 were carried out at the
ESO 3.6m telescope in December 1995 and at the CHFT in March 1997.  At ESO, we
used the Faint Object Spectrograph and Camera with the grism O300, yielding a
dispersion of 230\AA/mm, and the TEK512 CCD chip ($27 \mu $m 512 x 512 pixels);
at CFHT we used the Multi Object Spectrograph facilities (Le F\`evre et al.
1994) with the grating O300, resulting in a dispersion of 240\AA/mm, and the
STIS2 CCD ($21 \mu $m 2048 x 2048 pixels).  These combinations of gratings and
detectors result in dispersions of 6.3\AA/pixel and 5\AA/pixel, respectively .
Typically, two exposures, of 2700s each, were taken for fields across the
cluster.  Wavelength calibration was done using arc lamps before each exposure
(Helium--Argon, and Helium--Neon lamps).  Data reduction was carried out with
IRAF \footnote {IRAF is distributed by the National Optical Astronomy
Observatories, which are operated by the Association of Universities for
Research in Astronomy, Inc., under cooperative agreement with the National
Science Foundation} , using the MULTIRED package (Le F\`evre et al. 1995).
Radial velocities were determined using the cross--correlation technique (Tonry
and Davis 1981) implemented in the RVSAO package (developed at the Smithsonian
Astrophysical Observatory) with radial velocities standards obtained from
observations of late-type stars and previously well-studied galaxies.
 
\begin{figure*}
\epsfxsize=15.0cm \epsfverbosetrue \epsfbox{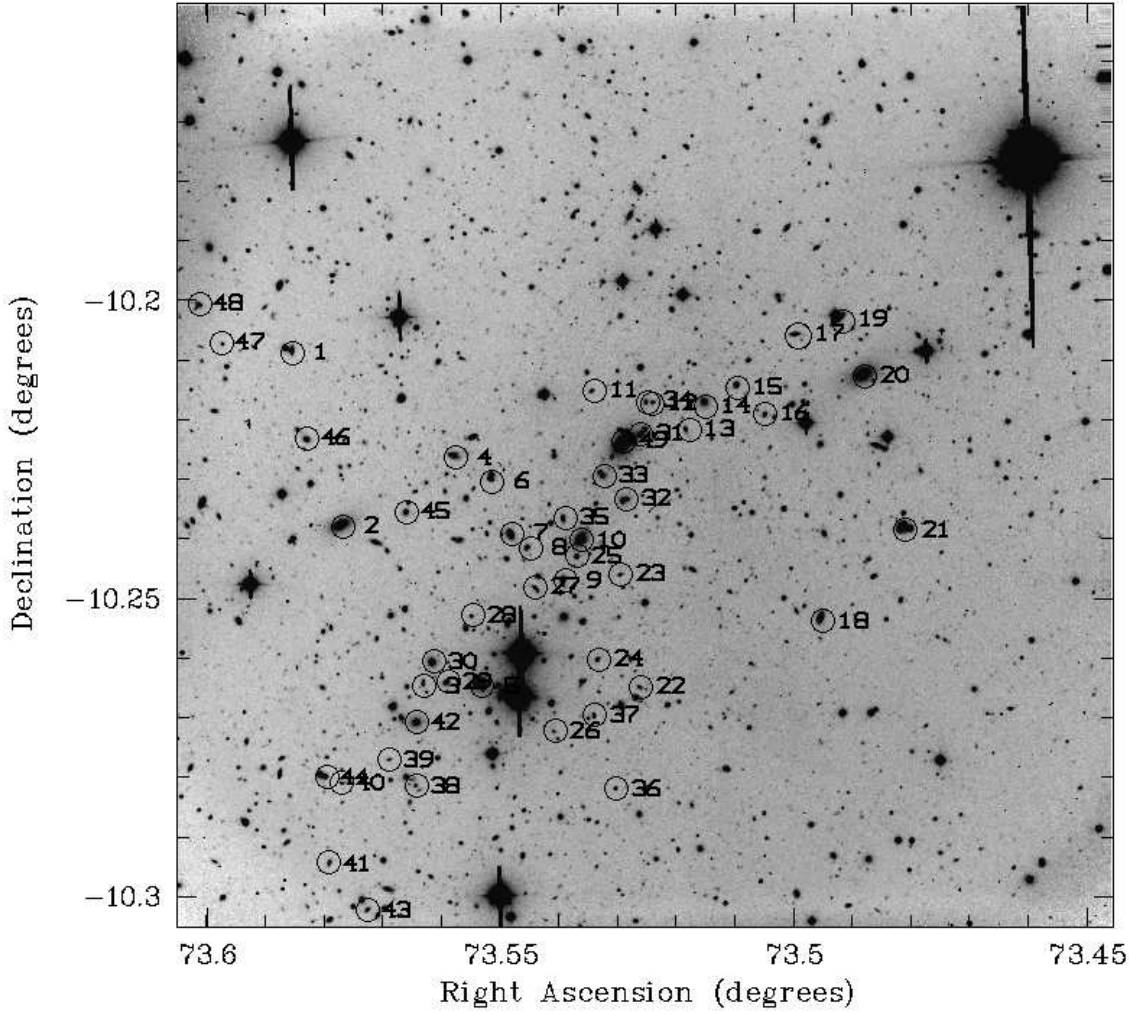}
\caption[]{Finding chart for galaxies with successful velocity measurements in
Abell 521. The galaxies are labeled as in Table 1.}
\end{figure*}
 
We have obtained 65 spectra of objects in the region of Abell 521. Star
contamination was very low (only 3 of the 65 targets turned out to be stars).
From these data we have retained 49 spectra of galaxies with a signal-to-noise
ratio sufficiently high to derive the measurement of the radial velocity with
good confidence (with the parameter R parameter of Tonry and Davis greater than
3). This results in a completeness of our velocity sample of 30\% of the
galaxies brighter than $m_V = 21$.  The finding chart for the objects with
measured velocities is shown in Figure 1.  Spectra of the brightest BCG
components were obtsained with EFOSC at ESO using the long-slit mode, and 3
different exposures ($2 \times 2700$ sec and 1800 sec).  Imaging of Abell 521
was obtained in 1994 at the CFHT with the MOS in B and R (using the LORAL3 CCD)
and in 1997 in the V and I bands (using the STIS2 CCD); the definition and
analysis of the galaxy catalog from the V-band image are presented in Arnaud et al.
1999 and the  complete photometric catalog will be provided in  Slezak et al.
1999. Also a J image of the cluster core was obtained in 
November 1996 at the CFHT with OSIS and 1800 sec of 
equivalent exposure (see Pell\'o et al. 1999).
Velocity measurements from our spectroscopy are listed in Table 1.  The
columns are as follows: column~(1): Identification number of each target galaxy
in the cluster as shown in the finding charts; Columns~(2) and~(3): Right
ascension and Declination (J2000.0) of the target galaxy; Column~(4): Best
estimate of the radial velocity resulting from the cross--correlation
technique; Column~(5): estimated error; Column~(6): a listing of detected
emission lines.  Galaxies 1-21 were observed at CFHT, while those from 22-49
were obtained at ESO.
 
\section{Velocity distribution in Abell 521}
\subsection{Global mean velocity and velocity dispersion of the cluster}
 
From a  visual inspection of the cone diagram displayed in Figure 2, we
selected a reasonable range of velocities (70000 to 80000 km/s) as candidate
members of the cluster. One galaxy (number 2) is identified as a clear
foreground object. Three background objects are found at a redshift $z \sim
0.295$; others are at redshifts of $0.289$, $0.31$ $0.331$ and $0.36$.
 
We have employed the ROSTAT package (Beers, Flynn, and Gebhardt 1990, hereafter
BFG) to analyze the velocity distribution of the 41 remaining galaxies in the
selected velocity range.  In order to quantify the central location and scale
of the velocity distribution for Abell 521, we have used the resistant and
robust biweight estimators ($C_{BI}$ and $S_{BI}$, respectively) recommended by
BFG.  For the complete sample of velocities, we obtain $C_{BI}= 74132 _{-250}
^{+202}$ km/s and $S_{BI}= 1386 _{-139} ^{+206}$ km/s .  Values obtained with
other estimators show similar values.  The one-sigma errors in these quantities
are calculated in ROSTAT by bootstrap re-sampling of 1000 subsamples of the
velocity data.
 
\begin{figure*}
\epsfxsize=14.0cm \epsfverbosetrue \epsfbox{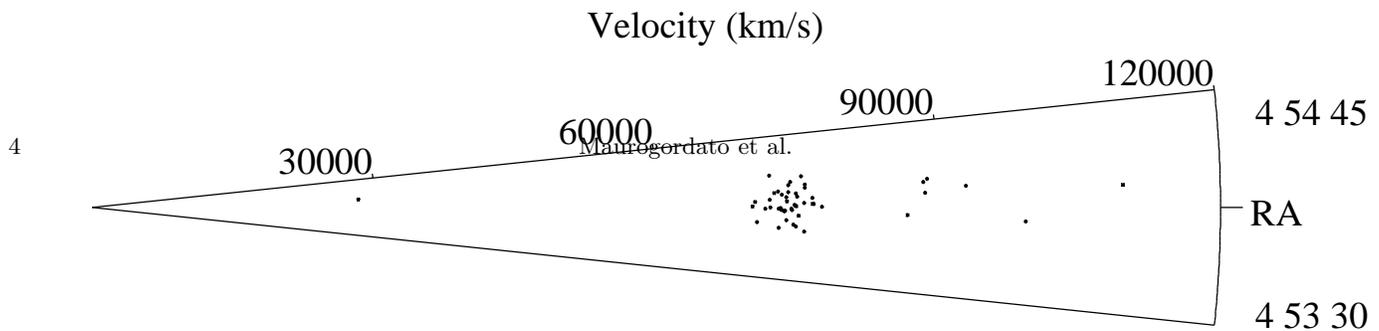}
\caption[]{Distribution in Right Ascension versus radial velocity for 49
galaxies in the inner $10^\prime \times 10^\prime$ region of Abell 521.}
\end{figure*}
 
In Figure 3 (top) we show a stripe density plot of the velocity distribution for
Abell 521.  The velocity histogram, calculated with a binning of $1000$ km/s,
is shown in Figure 3 (bottom), along with a superposed Gaussian of standard deviation
$1386 $km/s, shifted to the velocity of the cluster.  The radial velocity of
the brightest cluster galaxy (hereafter) BCG is shown with an arrow.
 
\begin{figure}
\epsfxsize=8.0cm \epsfverbosetrue \epsfbox{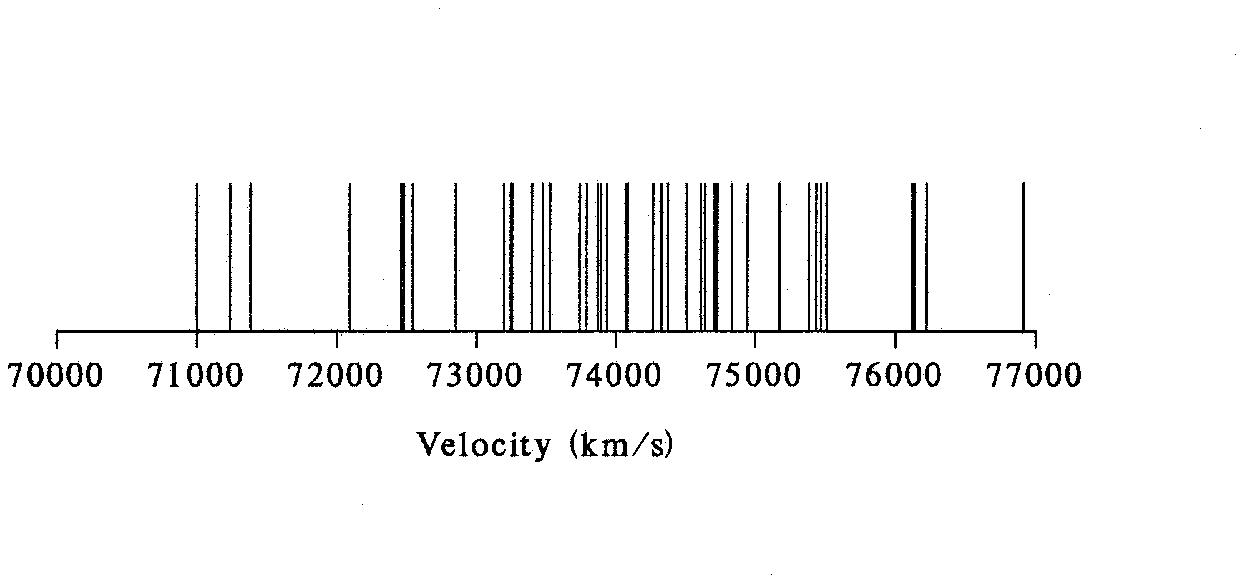}
\epsfxsize=8.0cm \epsfverbosetrue \epsfbox{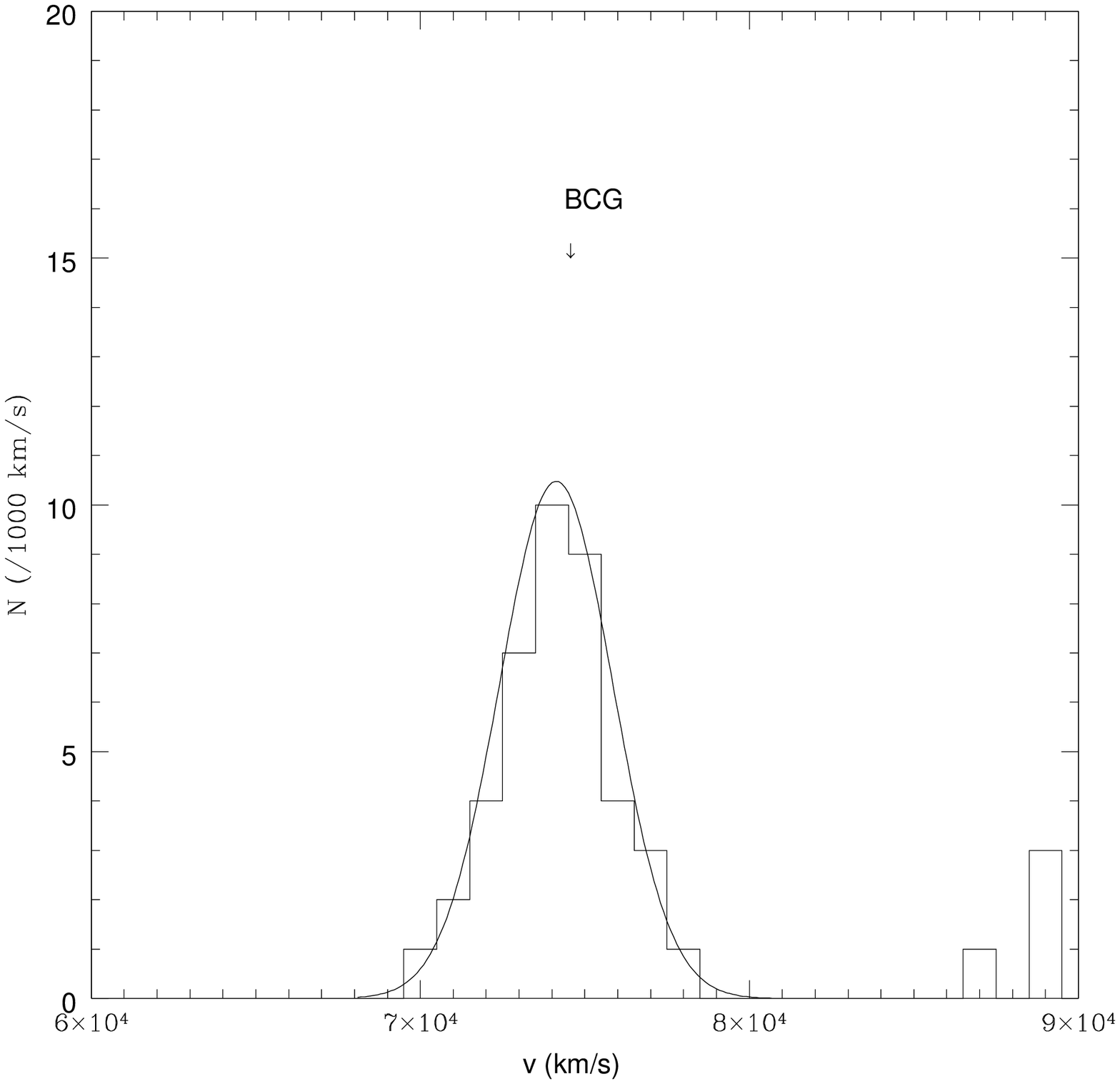}
\caption[]{(Top) Stripe density plot of radial velocities for presumed members of
Abell 521. The velocity of the BCG is shown with an arrow.  (Bottom) The radial
velocity histogram, calculated with a binning of $1000$ km/s, in a 10' x 10'
region centered on Abell 521.  A Gaussian of standard deviation $1386 $km/s,
shifted to the velocity of the cluster, is superimposed.}
\end{figure}
 
The apparent velocity dispersion of Abell 521, $\sigma \sim 1400$ km/s, is
among the largest values observed in galaxy clusters, as compared, for example
to the velocity dispersion distribution of the recent ENACS survey (Mazure et
al. 1996), and it is well above the median value of 744 km/s estimated by
Zabludoff et al. (1990) for a sample of 65 clusters.  However, the velocity
dispersion of Abell 521 is significantly {\it larger} than the value ($\sigma =
1017 \pm 65 $ km/s) we would predict from X--ray observations, using our
measurement of the gas temperature, and assuming equipartition between
kinetical and potential energy (Arnaud et al. 1999).
 
We endeavor to determine how reliable this estimate of the velocity dispersion
is, and whether or not it is affected by various problems such as subclustering
or contamination by outliers.  A fair assessment of the impact of potential
interlopers is essential for derivation of an unbiased measurement of the
velocity dispersion (see for instance Mazure et al. 1996).  Given that the
spatial coverage of our velocity sample is far from complete, and strongly
favors the high-density regions, and the limitations imposed by the relatively
small number of measured velocities, we cannot proceed to a sophisticated
analysis of subclustering.  For the present, we limit ourselves to classical
tests which examine whether our velocity measurements are drawn from a single
parent population, or are drawn from a mix of slightly-offset velocity
distributions which, taken as a single kinematic entity, would mimic this large
velocity dispersion.
 
\subsection{Simple statistical tests of the velocity distribution}
 
We have performed a number of statistical tests of the velocity data for Abell
521.  All twelve of tests implemented in ROSTAT are
consistent with the hypothesis that the velocities are drawn from a Gaussian
parent population.  We also searched for the existence of statistically
significant gaps in the velocity distribution, which can indicate the possible
presence of subclustering, especially when located in the center of a
distribution -- none were found. Bird and Beers (1993) discuss alternative
measures of the classical coefficients of skewness and kurtosis, the asymmetry
($AI$) and tail indices ($TI$), which are useful for detecting subtle
deviations from normality in distributions.  For the complete velocity set, we
obtain $AI = -0.238$ and $TI = 1.165$, respectively.  Neither of these values
allow rejection of a Gaussian parent population according to the tables
supplied by Bird and Beers.
 
While we cannot exclude some contamination from outliers or groups along the
line-of-sight to Abell 521, these results do exclude the presence of
significant projection effects in velocity space.
 
\subsection{Testing for substructure in the projected and redshift distribution}
 
To search for the presence of substructure in the projected galaxy distribution
of Abell 521 we have fit the observed galaxy positions to a number of Gaussian
mixture models, following the procedures described in Kriessler and Beers
(1997).  In this analysis we have only used the $\sim 400$ galaxies brighter
than $m_V=22.0$, to limit contamination from background galaxies projected on
the face of the cluster.  Figure 4 shows an adaptive-kernel contour map of the
region centered on Abell 521.
 
\begin{figure}
\epsfxsize=6.5cm \epsfverbosetrue \epsfbox{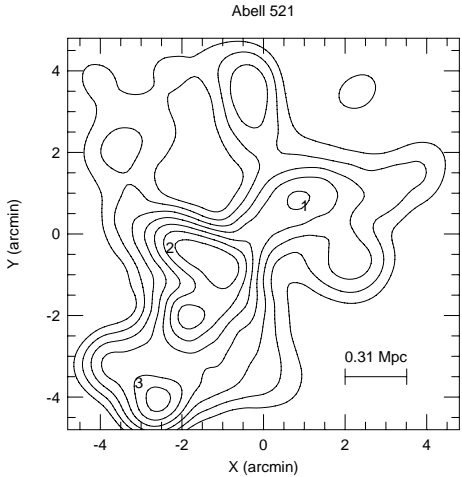}
\caption[]{Adaptative kernel contour map of the projected 
galaxy distribution
for  galaxies brighter than $m_V=22.0$. The centers of the three 
subclusters
corresponding to the KMM best partition are numbered as in Table 2.  The
minimum contour corresponds to a level of 3.638 galaxies/arcmin$^2$; the
contours are linearly spaced with a separation of 1.073 galaxies/arcmin$^2$.}
\end{figure}
 
\begin{figure}
\epsfxsize=6.5cm \epsfverbosetrue \epsfbox{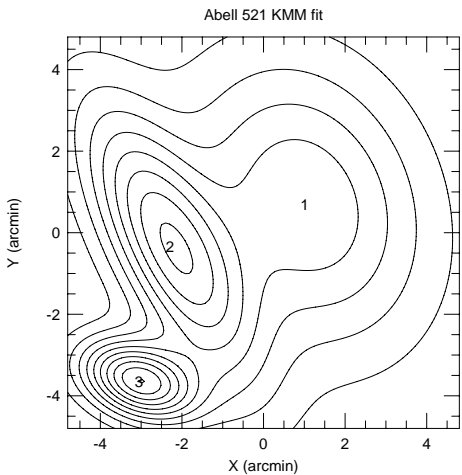}
\caption[]{Reconstructed contour maps of the three best-fit two-dimensional
Gaussians obtained from the KMM analysis (Table 2).  The minimum contour
corresponds to a level of 0.130 galaxies/arcmin$^2$; the contours are linearly
spaced with a separation of 1.329 galaxies/arcmin$^2$.}
\end{figure}
 
The best-fit KMM partition of the projected galaxy positions, evaluated using a
maximum-likelihood ratio test and a bootstrap procedure, is a three-group
partition which is significant at the 99\% level (parameters of the partition are
specified in Table 2).  Column (1) of this table lists the identification
number of the group (indicated in Figure 4).  Column (2) lists the number of
galaxies assigned to each group.  Columns (3) and (4) list the fraction of the
total number of galaxies present in each group, and the fraction of total
luminosity in each group, respectively.  The $x$ and $y$ positions of the
groups, along with their one-sigma errors, are listed in columns (5) and (6).
The median magnitude of the galaxies within each group is listed in column (7);
column (8) lists the mean magnitude of the 10th to 20th brightest galaxies in
each group.
 
From the application of a K-S test to the magnitude distributions of the
various groups, we find that group 3 is marginally fainter than the others.
This, along with the fainter value of $m_{10-20}$ (see Jones and Mazure 1996),
suggests that group 3 may contain a large fraction of background galaxies.
Figure 5 shows the reconstructed contour maps  of the three Gaussians
corresponding to the best-fit partition obtained with the KMM algorithm.
 
\begin{figure} 
\epsfxsize=8.0cm \epsfverbosetrue \epsfbox{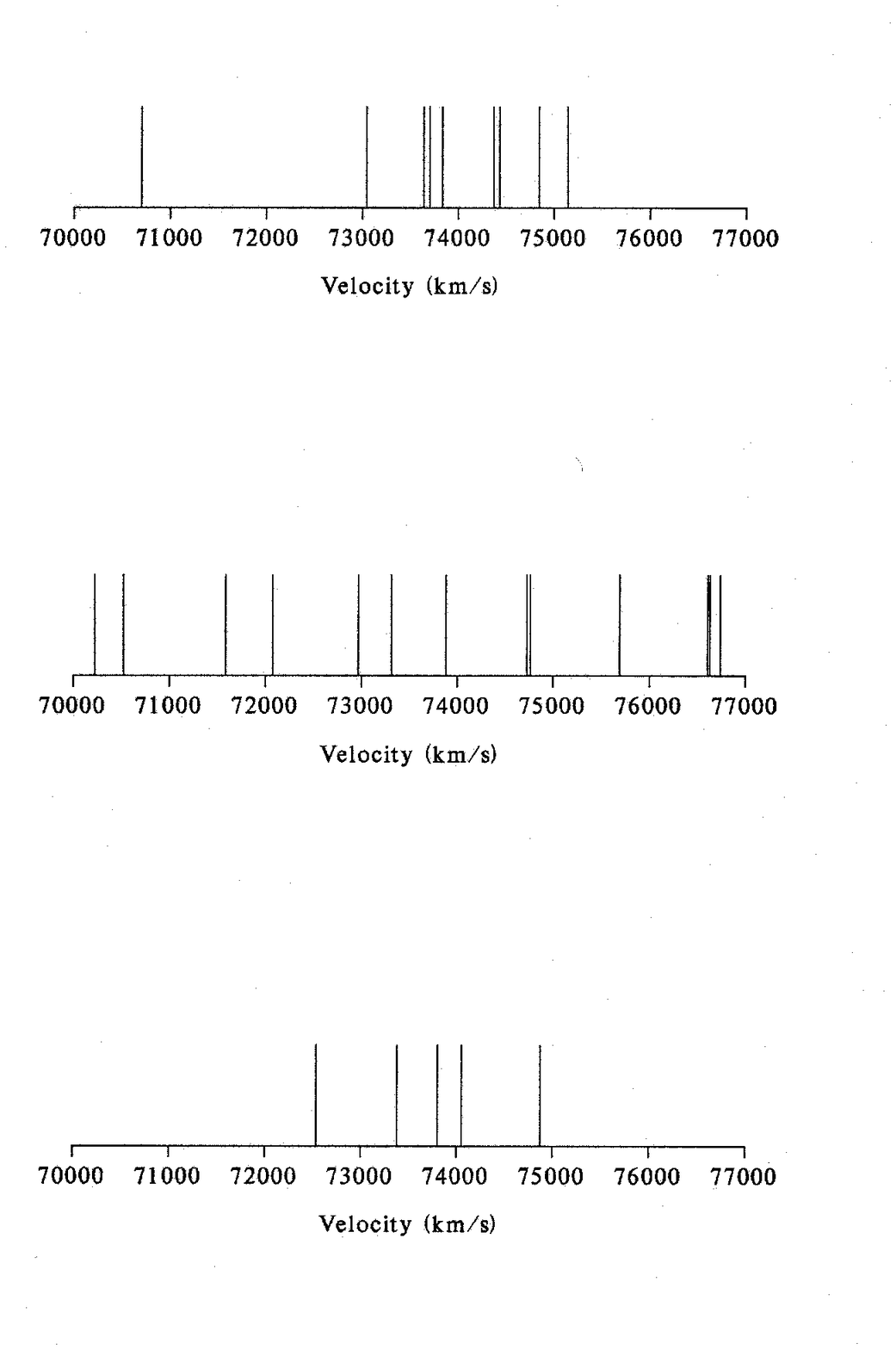}
\caption[]{Stripe density plots of radial velocities for the three
partitions analysed with ROSTAT -- from top to bottom: KMM1 ,  KMM2 North, and 
KMM2 South.}
\end{figure}
 
We next obtain a split of the velocity sample, assigning each galaxy to
a group associated to the nearest projected group center obtained from the KMM
analysis.  This results in 9 galaxies associated with KMM1, 19 with KMM2, and
2 galaxies with KMM3.  Galaxies located farther than 1.5 arcmin from any
of the group centers are set aside.
 
We then obtain a further split of the KMM2 group into two components:  KMM2
North (14 galaxies) and KMM2 South (5 galaxies) in order to isolate the
Southern extension of KMM2 seen in the adaptive kernel map shown in Figure 4.
There are only two galaxies with measured velocities assigned to the KMM3
region (numbers 40 and 42), both of which have slightly higher velocities
than the adopted central location velocity for the cluster as a whole.
More measured velocities are required to reliably determine the
the mean velocity of KMM3.
 
We are thus left with three subsamples of the velocity catalog on which we
have run the ROSTAT package, corresponding to regions KMM1, KMM2 North, and
KMM2 South.  The results of this analysis are summarized in Table 3.  Although
the small number of velocities in each subsample do not allow us to derive
precise measurements of the velocity dispersion, two qualitative conclusions
can be drawn.  First, there are no significant velocity offsets between the
individual partitions with respect to one another, at least to within the
bootstrapped errors on the velocity locations.  Second, the KMM2 North
partition has a significantly higher value of the velocity dispersion than the
other two partitions, or as compared to the cluster as a whole.  This result is
also strikingly clear on the stripe density plots of these three partitions
displayed in Figure 6.  The KMM2 North group, which includes the
so-called ``ridge'' structure described by Arnaud et al. (1999), is probably
kinematically complex, and may well be comprised of several subclusters.
 
We have also examined the location of the three galaxies with $z
\sim 0.295$ (objects 4, 40, and 46) to check if they are spatially clustered,
as would be expected for a background group.  These galaxies are all located in
the Eastern part of the cluster. Taking into account the velocity measurement
in Table 1, galaxies 4 and 46 are separated from one another by $\sim 3.2
h_{50}^{-1} $Mpc , and lie at the Eastern part of the KMM2 structure.  Galaxy
40 lies $\sim 2.4 h_{50}^{-1} $ Mpc apart galaxy 4, and $\sim 5.2  h_{50} ^{-1}
$ Mpc apart galaxy 46  in the Southern direction. It is thus not excluded that
these three galaxies are members of a background loose group, but a much more
complete redshift survey of Abell 521 is required to to resolve this question.
 
\subsection{Analysis of the velocity distribution with the color index}
 
Several analyses have shown that the velocity distribution of galaxies in
clusters can be very different for individual morphological types (e.g., see
the analysis by Binggelli et al. 1987 on the Virgo Cluster, Beers et al.  1992
on A400, Bird et al. 1995 on Abell 151, and of Girardi et al. 1996 on a larger
sample of clusters).  A higher value of velocity dispersion is generally found
for late-type galaxies, as compared to early types, which is expected if the
latter have fallen into the cluster potential following the initial collapse
(Tully and Shaya 1984).  The spatial resolution of our imaging data for Abell
521 is unfortunately not sufficient to assign a morphological type to all the
objects with measured velocities, in particular at the faintest magnitudes.  As
an alternative, we have used the color indices of the galaxies in order to
define two subsamples within our velocity catalog, with values of $B-R$
respectively higher and lower that the median value for the sample as a whole.
In the following we refer to these as the ``{\it red}'' and ``{\it blue}''
subsamples.
 
Inspection of the brightest galaxies, whose morphological type is unequivocally
determined by eye, shows that typical cluster ellipticals belong to the {\it
red} subsample, and  spirals to the {\it blue} subsample.  Galaxies of the
cluster with detected emission lines (Table 1) belong to the {\it blue}
subsample, as expected.  Figure 7 shows stripe density plots of the velocity
distributions for the {\it red} and {\it blue} subsamples.  These subsets
appear rather different.  ROSTAT analysis of the two subsamples yields values
of $C_{BI}=74125 ^{+218}_{-273}$ km/s and $S_{BI} = 1011 ^{+214}_{-108}$ km/s
for the {\it red} subsample of 19 galaxies, and $C_{BI}= 73924 ^{+408}_{-445}$
km/s and $S_{BI} = 1803 ^{+256}_{-191}$ km/s for the {\it blue} subsample of 20
galaxies, respectively.  The central locations on velocity of the high and low
$B-R$ subsamples are consistent with one another, but the velocity scales are
significantly different.  We have further checked how stable these results are
by testing different subsamples obtained by translating the color cut on $B-R$
within $1\sigma$ around the median value.  The range is of course limitated, as
the number of objects falls rapidly when moving off the median.  The values of
the dispersions which are obtained fluctuate $\sim 20 \%$ around the previously
calculated ones, but the general trend remains true and becomes even more more
accentuated when taking only the most extreme regions of the color
distribution.  In any case, the velocity distribution of the {\it blue}
subsample is quite large, while the {\it red} subsample shows a value which is
typical of other galaxy clusters.
 
One might wonder if the colors of the galaxies are correlated with the
high-density structures evidenced in the V-band density map.  Based on our
inspection of the color indexes, there might be a color segregation in the
various structures, with a bluer NE/SW extension including KMM2 North and more
compact redder clumps along the NW/SE extension.  However the bluest galaxies
belonging to our velocity sample are distributed across the entire field of the
cluster, so the high velocity dispersion of these galaxies cannot be due solely
to the contribution of the KMM2 North region.
 
The higher velocity dispersion of the {\it blue} subsample can be explained
under the hypothesis that Abell 521 is in fact dynamically complex, and one
might expect this class of galaxies includes many spirals which are not yet
virialized within  the cluster potential.  In this case the distribution of
spirals would be much more dispersed than that of ellipticals (Girardi et al.
1996).  In fact, the velocity dispersion as estimated from the {\it red}
subsample, $S_{BI} = 1011 ^{+214}_{-108}$ km/s, is quite in line with the
predicted cluster dispersion based on the X--ray analysis, which is at variance
with the dispersion based on the entire set of galaxies with measured
velocities.  
 
\begin{figure}
\epsfxsize=8.0cm \epsfverbosetrue \epsfbox{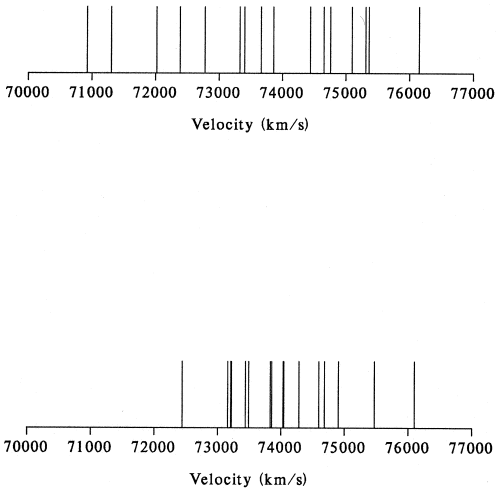}
\caption[]{Stripe density plots of radial velocities for the two subsamples
split according color index:  (Top) the {\it blue} subsample,
and (Bottom) the {\it red} subsample.}
\end{figure}
 
\section{The region surrounding the brightest cluster member}
 
\subsection{The brightest cluster galaxies in Abell 521}
 
In Figure 8 we show a one arcmin subframe of the 600s R image centered on the
BCG in Abell 521. This galaxy has a magnitude $m_V=17.22$, which corresponds to
a K-corrected absolute magnitude $M_V= -24.37 + 5\log h_{50}$. Two other gE
galaxies are present in the core of Abell 521, identified in Figure 1 as
objects 10 and 20 with magnitudes 18.71 ($M_V= -22.90 + 5 \log h_{50}$) and
18.67 ($M_V= -22.86 + 5 \log h_{50}$), and velocities of 74763 km/s and 73044
km/s, respectively. The K-correction for E/SO galaxies is computed using
 the spectral
energy distribution of an old elliptical galaxy and the filter
response. The synthetic spectrum was obtained through the GISSEL96
evolutionary code (Bruzual \& Charlot 1993), with the Miller
\& Scalo IMF (1979) and solar metallicity.
 In their analysis of a statistical sample of 116 nearby
Abell clusters, Hoessel et al.  (1980) found an average value of $M_{V}=-22.68
\pm 0.03$ for the cluster BCG's with a dispersion of 0.35 magnitudes within a
$16.4$ kpc aperture. We have calculated the K-corrected absolute magnitude in
this same aperture (4 arcsec at the redshift of Abell 521) and using the same
cosmological parameters ($H_0 = 60$ km/s/Mpc), and we obtained respectively:
$M_V= -22.57$, $M_V= -22.28$ and $M_V= -22.37$ for the BCG, galaxy 10, and galaxy 20, which are all within $1.5 \sigma$ of their findings.  Thus Abell 521 includes, besides
the BCG, two other giant elliptical galaxies of central absolute magnitude
typical of brighter cluster members.  These objects could be interpretated, in
the hierarchical scenario in which rich clusters form from the aggregation of
smaller units, as the main galaxies of the groups which have collided. The
presence of these giant ellipticals, which are expected to eventually accrete
to the most massive giant elliptical, would then suggest that we are seeing the
early stage of the formation of the cluster.
 
The BCG in Abell 521 extends over at least  20 x 30 arcsec , which corresponds
to spatial dimensions of 90 x 150 $h_{50}^{-1}$ kpc.  This is why, although the
BCG luminosity obtained within the previous small apertures is comparable to
the other two giant ellipticals, its asymptotic total luminosity is
significantly larger.  In fact, its high absolute luminosity and large
dimensions are quite characteristic of a cD galaxy.  However, the original
definition of a cD galaxy (Matthews et al. 1964) also requires that it be embedded in
an extended luminous envelope which is generally detected as a flattening of
the slope in the surface brightness profile at large radius (see also Schombert
1988).  Unfortunately, our image is neither deep enough nor large enough to
detect the existence of such a halo.  We are therefore unable, with the present
data, to disentangle between the cD/D/gE classifications for the BCG.  If the
BCG is a cD galaxy, then in this analysis we are missing the contribution of
the halo to the total luminosity, and our  measured magnitude is probably
under-estimated.
 
\begin{figure}
\epsfxsize=8.0cm \epsfverbosetrue \epsfbox{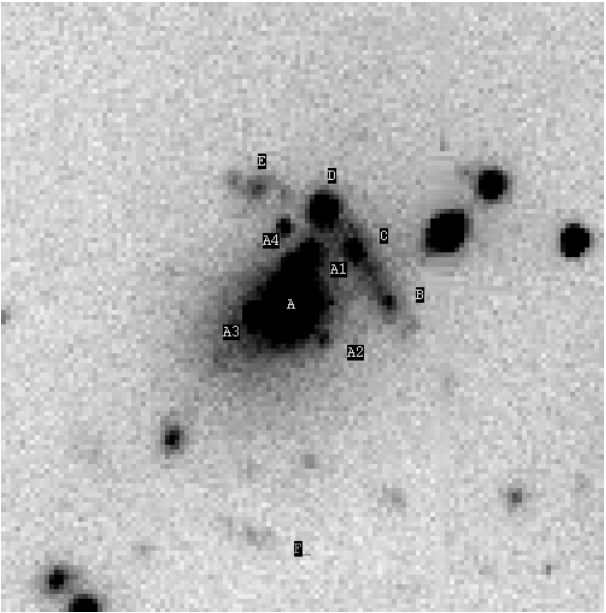}
\caption[]{Close-up on the region of the Brightest Cluster Galaxy.  Multiple
nuclei are apparent near the central galaxy A (e.g., A1, A2, A3, and A4).  The
giant arc candidate is visible to the northwest at a distance of $\sim 7$
arcsec from the BCG galaxy as a diffuse structure with brighter knots
superimposed (B, C, D, and E).  A smaller arc candidate (F) is visible at 14
arcsec to the South.}
\end{figure}
 
\subsection{The core of the BCG}
 
The very central region around the BCG galaxy shows a complex structure.  In
the inner region of $ 20 h_{50}^{-1} kpc $, one can notice a bright secondary
nucleus, A1, and three faint ones: A2, A3, and A4 (see Figure 8).  Multiple
nuclei in brightest cluster galaxies have shown to be quite a frequent
phenomena.  Various systematical studies of samples of BCGs in clusters (Tonry
1985; Hoessel and Schneider 1985) have shown that about $50 \%$ of BCGs have at
least one extra nucleus within $20 h_{50}^{-1} $kpc.  Only a small fraction of
these occurances have been shown to be attributed to spurious projections.  We
successfully measured the redshift of nucleus A1 ($73857 \pm 55 km/s$),
confirming its status as a cluster member.  Although we did not measure the
redshift of the fainter nuclei A2, A3, and A4, their color indexes are rather
similar to those of A1, and are thus expected to be members of the cluster as
well.  The velocity measurement of nucleus A1 shows a significant peculiar
motion with respect to the BCG of $-413 \pm 57 km/s$ in the local rest frame.
These results suggest that the region of the BCG is still experiencing strong
dynamical activity, and that, according to the scenario of Hausman and Ostriker
(1978), the multiple nuclei are the remnants of galaxies absorbed by the BCG by
cannibalism.
 
\subsection{The giant gravitational arc candidate revisited}
 
Roughly 7 arcsecs to the Northwest of the BCG, one can see the giant arc
candidate studied by Maurogordato et al. (1996) (Figure 8). It spreads over the
northwest quadrant more than 10 arcsecs, with a radius of curvature of $\sim
7.5$ arcsecs, and it is nearly centered on the BCG.  The structure is clumpy
and appears as a chain of several bright condensations (B, C, D, and E)
superimposed on a possible diffuse component.  Also, an arclet candidate
appears 14 arcsecs to the South (F). The mean magnitudes and colors for the
central galaxy and the different knots found in this area are listed in Table
4.  The colors of the brightest knot D are fully compatible with the expected
value for an elliptical galaxy at the cluster velocity, and it is also in good
agreement with the colors observed for the brightest galaxies in the cluster
core. The velocity of this object ($74379 \pm 72$ km/s) clearly identifies it
as a cluster member.  A moderate signal-to-noise spectrum of knot D is
presented in Figure 9, and compared to a synthetic spectrum of a E-type galaxy.
The condensations along the arc candidate display different colors in the
filter-bands available. This fact  a strongly argues against the hypothesis
of a single multiple-image system.
 
We would like now to compare the spectral energy distribution of the three
knots in detail. On this purpose, as no spectrophotometric standard star was
observed during our run, we have used the good-quality spectrum of knot D to
flux calibrate the other two components (Figure 9).  As knot D exhibits all the
spectral features expected for a typical elliptical galaxy, a synthetic E-type
continuum was used for this exercise, taken from the GISSEL evolutionary code
(Bruzual and Charlot 1993, updated in 1998).  Knots C and D have the same
redshift within the errors ($z=0.248$). The spectrum of knot B is noisier, but
its redshift is also compatible with that of the cluster ($z=0.25$, see Table
4).  The spectra of these knots become bluer from D to B, as expected from
their $B-R$ colors.  At the redshift of this cluster, the $B-R$ color index is
a good measure of the Balmer break, or the 4000 \AA \ break.  Furthermore, the
relative strength of the redder Balmer absorption lines increases from D to B.
Knot C exhibits absorption lines similar to those found in spiral galaxies at
such a resolution, whereas knot B displays much stronger Balmer lines.
Unfortunately, the poor quality of the knot B spectrum precludes a careful
synthesis of the stellar population.  Although a more careful analysis of these
objects is needed, based on spectra with better signal-to-noise ratios, these
preliminary results indicate a clear change in the properties of the stellar
population along the arc.  If one interprets the change in Balmer line
strengths as indicative of an age sequence, knot B is the younger, and the mean
age of the underlying stellar population increases towards knot D.
 
According to these results, the three brightest knots of the arc structure seen
in Abell 521 are in fact cluster galaxies. This result does not completely
exclude the gravitational lensing hypothesis for the other arclet candidates in
this region (E,F).  A deep photometric multicolor survey is required for
further progress, in particular to identify possible background sources using photometric redshift techniques.
 
\begin{figure*}
\vbox{
\hbox{
\psfig{figure=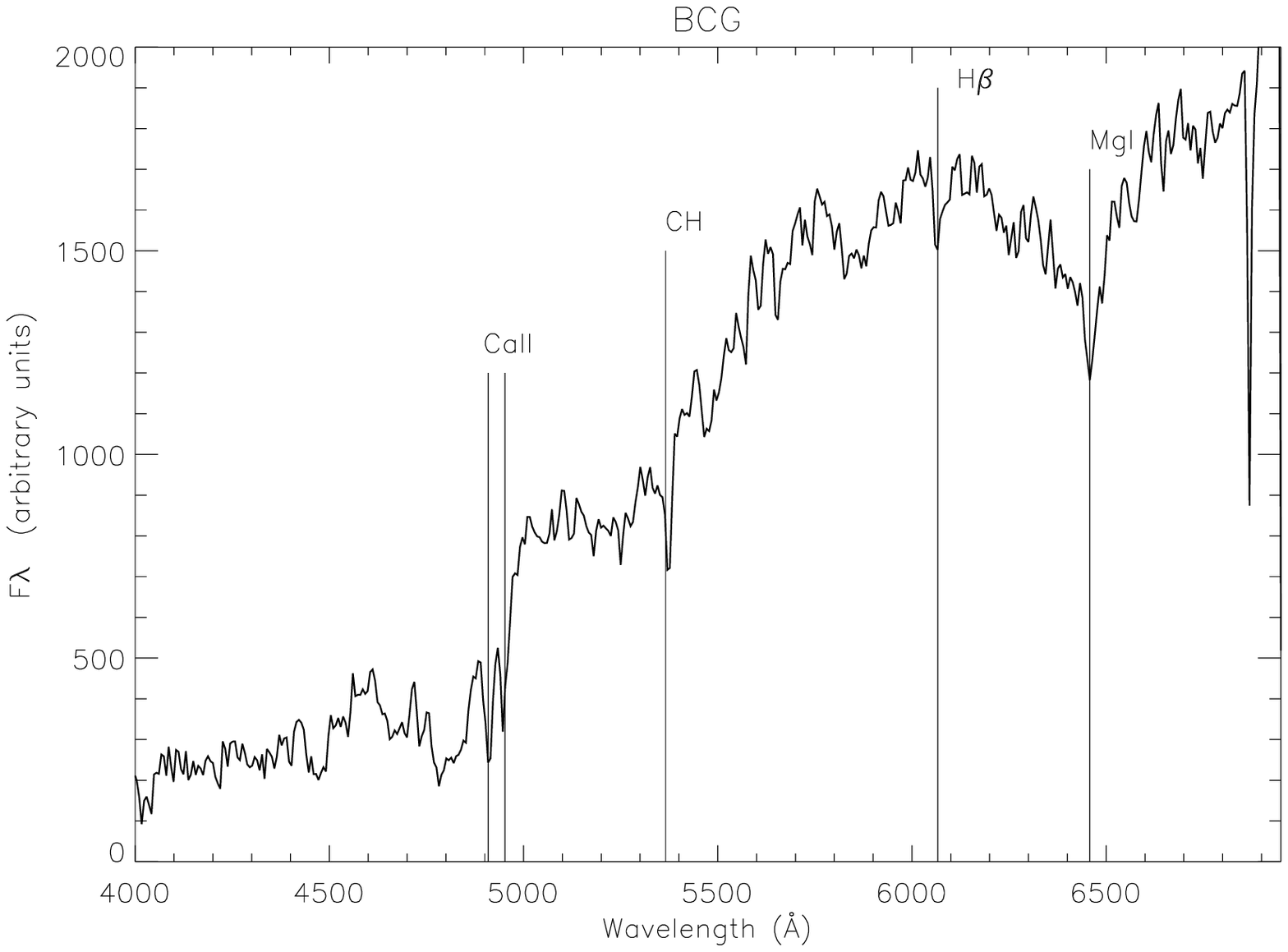,height=6.5cm}
\psfig{figure=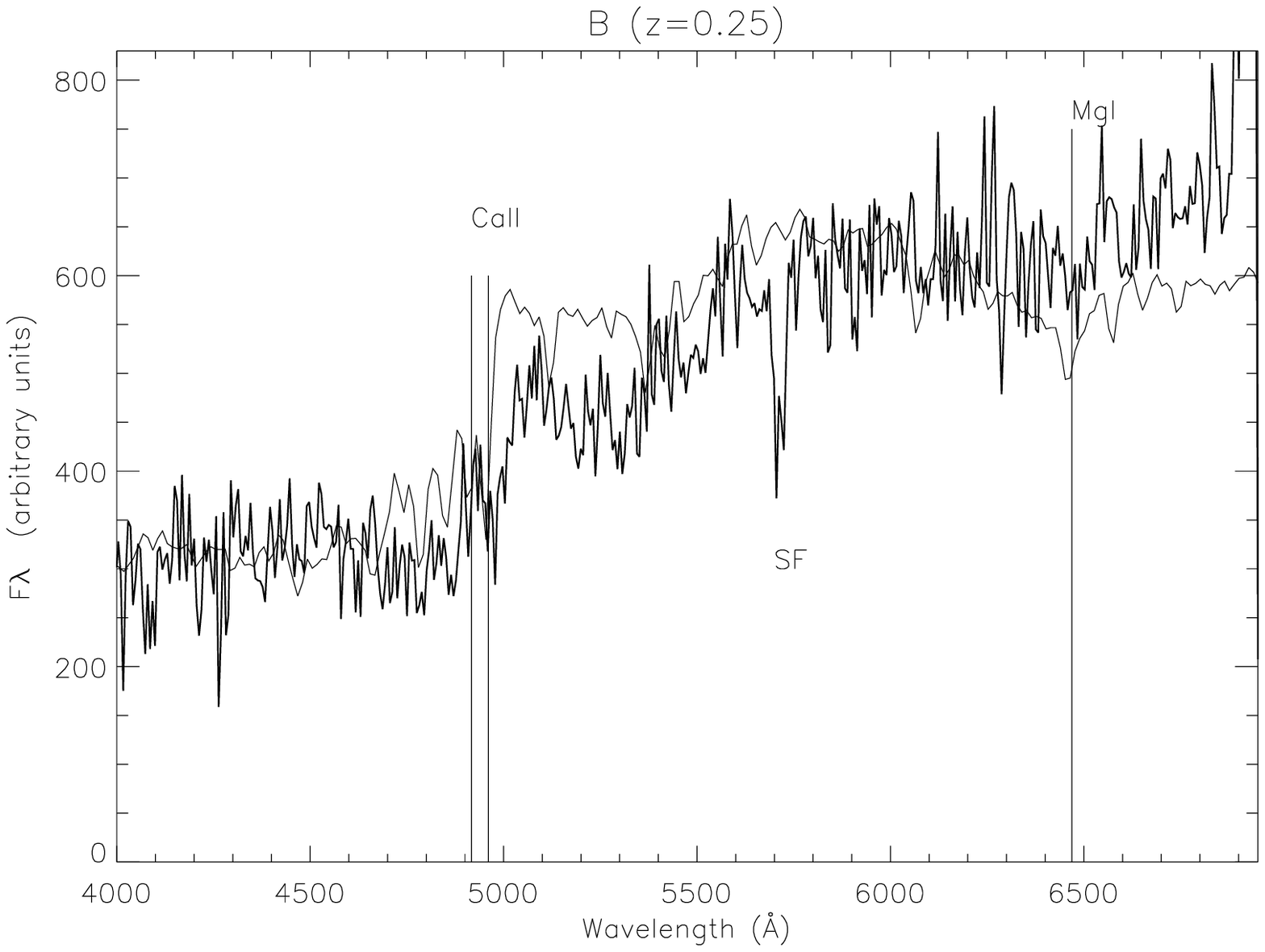,height=6.5cm}
}
\hbox{
\psfig{figure=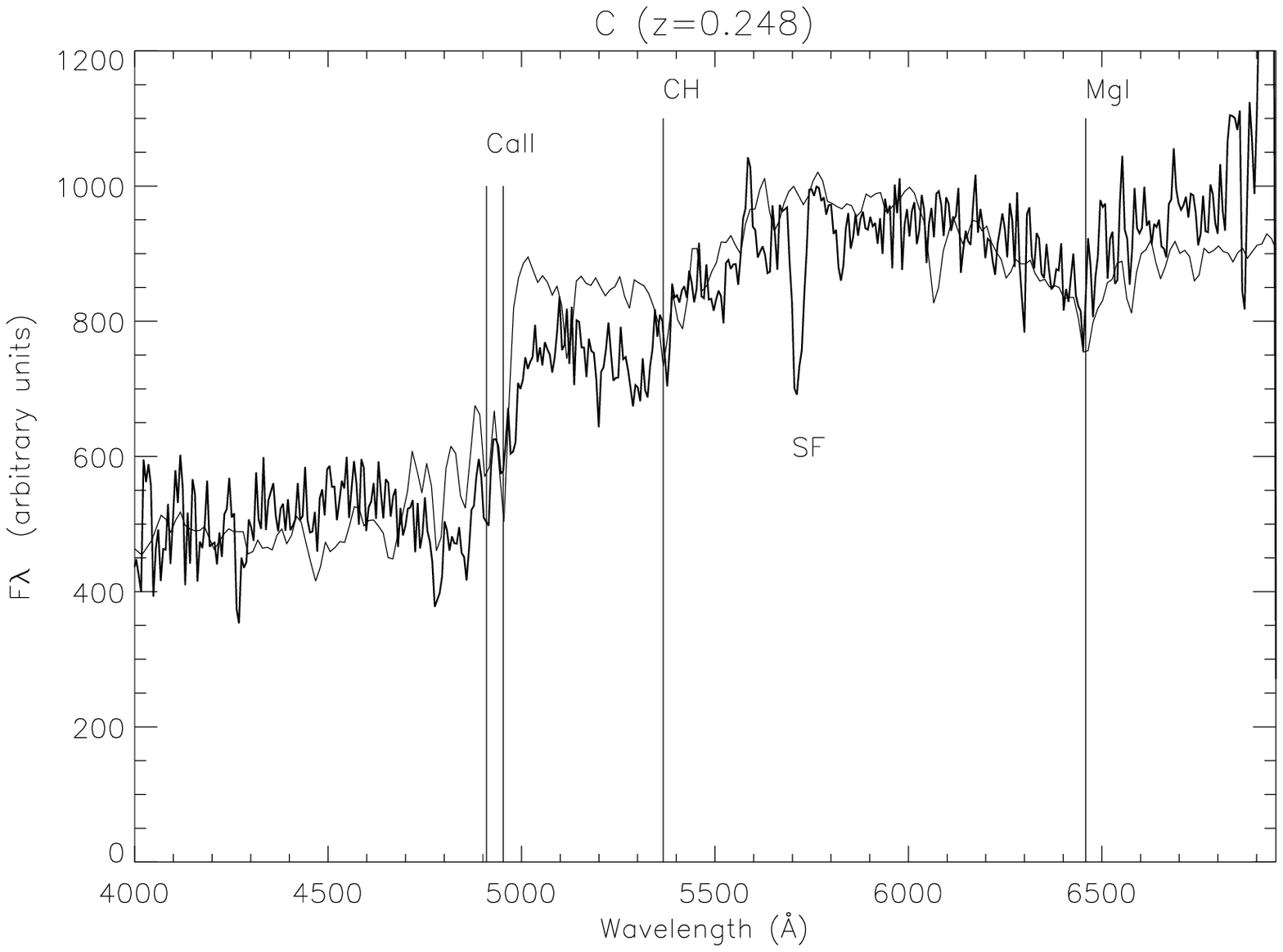,height=6.5cm}
\psfig{figure=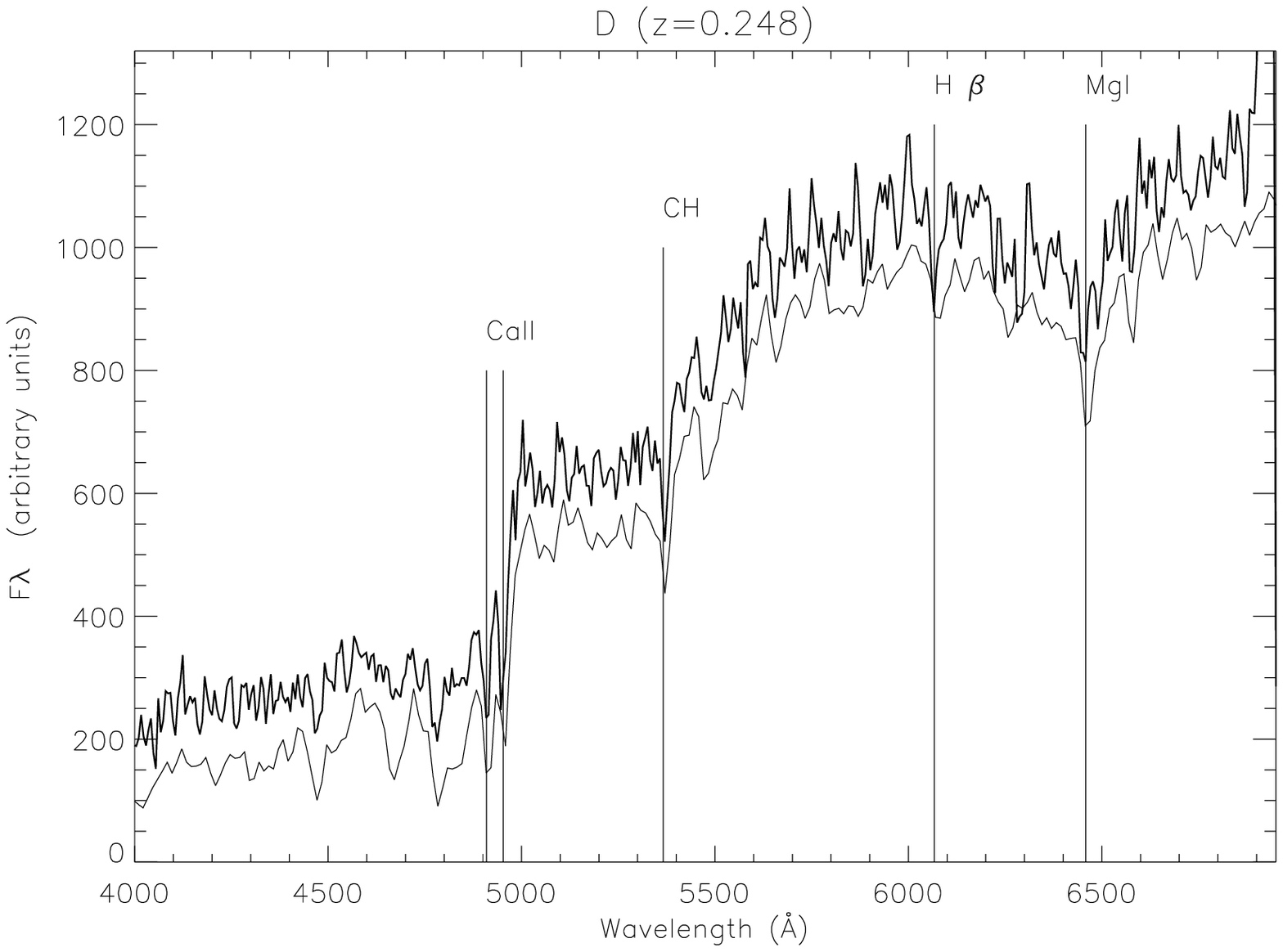,height=6.5cm}
}
\caption{Spectra of the BCG (Top-left), and of the various bright
components of the arc: knot B (Top-right), knot C (Bottom-left), 
Knot D (Bottom-right). A synthetic spectrum of an S-type
galaxy is superimposed on the spectrum of B and C for comparison. An
E-type synthetic spectrum is overplotted on the spectrum of the bright
knot D.}
}
\end{figure*}
 
\subsection{Peculiar motion of the BCG ?}
 
In most theories of BCG formation in clusters, these objects are required to
occupy the bottom of the cluster gravitational potential.  However, there have
been suggestions of evidence for at least some BCGs presenting significant
velocity offsets as compared to the velocity location of the cluster as a whole
(Oegerle and Hill 1994). It has been pointed out that many of the apparently
large peculiar velocities seem to be correlated with the presence of
substructure, and have been claimed to represent a signature of recent (or
ongoing) subcluster merger events (Bird 1994).  Moreover, at the bottom of the
cluster potential the effect of the gravitational redshift becomes
non--negligible (Nottale 1983); as shown by Cappi (1995), in rich clusters this
effect gives a significant positive velocity difference between the BCG and the
cluster and, while in principle detectable only statistically with a large
number of clusters, it represents an additional source of incertitude for the
estimate of the BCG peculiar velocity.
 
Care must be taken when evaluating the significance of a proposed BCG velocity
offset (see the discussion in Gebhardt and Beers 1991), taking appropriate
estimates of errors in central location on velocity into account.  As we are
confronted with observational evidence of merging activity in Abell 521, we are
particularly interested in quantifying whether a peculiar offset of the BCG
does in fact exist.  The BCG galaxy of Abell 521 has a velocity $v_{BCG} =
74374 \pm 47 $km/s, slightly  higher than the central velocity location of the
cluster as a whole $C_{BI} = 74132 ^{+202}_{-250}$ km/s.  This results in a
peculiar velocity $v_{pec} = +242$km/s, which has to be corrected by ($1+z$) to
obtain the peculiar motion in the cluster rest frame: $ v_{pec} = +194 $km/s.
Since we have argued above that the kinematics of Abell 521 may in fact be
rather complex, we have also calculated the velocity offset of the BCG with
respect to the {\it red} subset of velocities.  After correction to the cluster
rest frame, we obtain a similar peculiar velocity: $v_{pec} = +200$ km/s.
 
Are either of the above offsets statistically significant?  Gebhardt and Beers
(1991) outline a prescription for answering this question, based on a
comparison of the measured (rest frame) velocity of the BCG with respect to the
biweight estimator of velocity central location for the parent cluster or
subcluster, with bootstrapped (90\%) confidence intervals on the velocity
central location.  The 90\% confidence interval on velocity location for the
entire set of galaxies is [$-398,+327$] km/s; that for the {\it red} subsample
is [$-432,+371$].  The 90\% measurement error for the BCG velocity is $\pm 78$
km/s.  Thus, in both instances, the peculiar velocity of the BCG is completely
contained with the 90\% interval on central location in velocity space, and
neither offset can be considered significant.  This probe could be
substantially improved by obtaining a larger set of galaxies with measured
velocities in Abell 521, in order to shrink the confidence intervals on
location in velocity space.
 
We have also calculated the peculiar velocities of the two bright elliptical
galaxies (10 and 20) and obtain rest-frame peculiar velocities of $v_{pec} =
+506 $km/s and $ v_{pec} = -872 $km/s, respectively, when compared to the
cluster as a whole.  When compared to the {\it red} subsample, the peculiar
velocities are $v_{pec} = +512 $km/s and $ v_{pec} = -867 $km/s, respectively,
which are essentially the same.  The 90\% measurement errors for the observed
velocities of these galaxies are $\pm 88$ and $\pm 150$ km/s, respectively.
The velocity offset of galaxy 10 is thus judged to be statistically
significant, when compared to either the entire cluster or the {\it red}
subsample. The 90\% confidence interval on the measured velocity fails to
overlap with the 90\% confidence intervals on the central locations in either
instance.  An even greater significance is attached to the velocity offset of
galaxy 20.
 
\section{Discussion and Conclusions}
 
From the analysis of our data obtained through multi--object spectroscopy, we
have found that Abell 521 is a moderately distant cluster ($z=0.2467$) with an
apparently very large velocity dispersion ($S_{BI} = 1386$ km/s).  The velocity
distribution of cluster members is consistent with sampling from a parent
Gaussian population, and the high dispersion does not seem to result from
trivial superposition effects.  However, there do exist several hints that this
large apparent velocity dispersion may be due to a superposition of several
distinct populations of galaxies.  The projected distribution of the galaxy
positions in the cluster can be described with a mixture of three
two-dimensional Gaussians, a model which is significant at the 99\% level.
Analysis of the velocity distribution for these partitions shows that the
velocity dispersion is very high ($\sim 2000$ km/s) in the central North-East/
South-West ``ridge'' corresponding to KMM2 North, although the dispersion for
the smaller samples of galaxies associated with the KMM1 and KMM2 South
partitions are more typical of most rich clusters ($\sim 800$ km/s).  We also
find that the velocity distribution is rather different for subsets of the
galaxies selected according to color, with the bluest objects (spirals) showing
a high velocity dispersion ($\sim 1800$ km/s) and the redder objects
(ellipticals) exhibiting a much smaller dispersion ($\sim 1000$ km/s).  This is
another indication that this cluster is dynamically young, with its population
of spirals not yet relaxed to the cluster potential.  Moreover, the high value
of the velocity dispersion as compared to the one expected from the X-Ray
Temperature from the $\sigma / T$ relation suggests that this cluster is far
from dynamical equilibrium.
 
On the basis of our results, we can outline a tentative picture of the
dynamical state of Abell 521.  In the frame of hierarchical models, the cluster
formation process proceeds by merging of smaller units (Frenk et al. 1996).
Seve ral evidences that  merging processes are occurring in this cluster have
been su ggested in the previous analysis.
 
However, one can note that the various clumps evidenced on the NW/SE axis
(KMM1, KMM2 South, KMM3) are well-defined concentrations. The foreground/background hypothesis is quite
improbable, as redshift measurements have shown that all the groups contain
several cluster members (although in the case of KMM3 only two redshift
measurements exist).  We are then probably seeing the early phase of infall of
these various groups. According to numerical simulations (Schindler and
Bohringer 1993), we should expect that groups are strongly accelerated and
their velocity distributions diverge as the merger process proceeds.  However,
we do not detect any offset of the central velocity locations of the various
groups (except possibly KMM3, which could be at a slightly higher velocity of
$\sim 75800 km/s$ based on the two measured velocities).  This implies that
either we are seeing the very beginning of the merging and the clumps are still
close to at rest with respect to one another, or we are witnessing a more
advanced state but fail to detect the shift of the velocity distributions
because the collision axis happens to be mostly in the plane of the sky.
 
The apparently large velocity dispersion of Abell 521 is due in great part to
the contribution of the KMM2 North structure.  Its very high velocity
dispersion suggests that we are witnessing the collision epoch, at which point
numerical simulations show that the dispersion reaches its maximum value, which
can be twice the value after the cluster approaches dynamical equilibrium
(Schindler and Bohringer 1993). The present data could be explained in a
scenario whereby two (or more) subclusters have just collided along an axis
which is projected on the sky in the direction of KMM2 North, but with a
substantial component along the line of sight.  The large peculiar velocities
of the two bright elliptical galaxies of the cluster is an indication that they
probably originated in different initial subclusters, as has been seen in other
clusters (e.g., Abell 2255, Burns et al. 1995).  A detailed merging sscenario
taking into account the whole set of optical and X-Ray properties of this
cluster is presented in Arnaud et al. 1999.
 
The region surrounding the Brightest Cluster Galaxy is particularly
interesting.  The inner part of the BCG ($20 h_{50}^{-1} kpc$) exhibits
multiple nuclei, one of which (A1) exhibits a significant velocity offset with
respect to the main body of the BCG (A).  These facts suggest the possibility
of strong merging activity.  At larger distance from the BCG ($30 h_{50}^{-1}
kpc$) a curved structure is seen, with three knots superimposed. Analysis of
the spectra and colors of the knots suggest that at least two of them are in
fact galaxies belonging to the cluster.  The spectrum of the third knot has a
low signal-to-noise ratio, but seems to also be at a redshift typical of the
cluster.  This calls into question the gravitational lensing hypothesis to
explain the arc-like structure, and favors the interpretation as an effect of
merging.  The colors of the two arclet candidates E and F suggests these should
be high-redshift objects.  Deep multiband observations are necessary to
constrain the redshift range of the knots by the technique of photometric
redshifts.  Additional multi-object spectroscopy in order to obtain a complete set of
velocity information for galaxies in Abell 521 down to magnitude $m_V=21$
 would be crucial
for our understanding of the dynamical state of this aparently highly
unrelaxed cluster.
\acknowledgements{ We would like to thank C. Vanderriest and O.Le F\`evre
for the acquisition of the first images of Abell 521 at CFHT, with Alain Mazure for
fruitful discussions on the dynamics of this cluster.}

\begin{table*}
\renewcommand{\arraystretch}{0.8}
\caption[]{Heliocentric redshifts in the 10'x10' frame centered on Abell 521.}
\begin{flushleft}
\begin{tabular}{llllll}
\hline
{\bf NUMBER}  & {\bf R.A.} & {\bf DEC.} & {\bf HEL. VEL.} &
{\bf ERROR} & {\bf emission lines}\\
       &   (2000)   &   (2000)   &    v (\kms)   &
${\Delta}$~v~(\kms) \\
\hline
1       &4:54:20.5 &-10:12:30.8 &74298        &84 \\
2       &4:54:18.4 &-10:14:15.9  &28324        &24  \\
3       &4:54:15.1 &-10:15:52.5 &72534        &78  \\
4       &4:54:13.8 &-10:13:34.2  &88586        &102 \\
5       &4:54:12.7 &-10:15:52.3 &73805        &38   \\
6       &4:54:12.3 &-10:13:48.4  &76607       & 57 \\
7       &4:54:11.5 &-10:14:20.5 &72071        &54 \\
8       &4:54:10.7 &-10:14:28.9 &73885        &73  \\
9       &4:54:09.3 &-10:14:48.4 &76630        &48  \\
10      &4:54:08.6 &-10:14:24.3  &74763        & 53 \\
11      &4:54:08.1 &-10:12:53.8  &74873        &67  \\
12      &4:54:05.7 &-10:13:01.4  &73640        &64 \\
13      &4:54:04.2 &-10:13:17.3 &86725        &206\\
14      &4:54:03.5 &-10:13:02.9  &75146       & 59  \\
15      &4:54:02.2 &-10:12:51.8 &99306        &33  &[OII],$H_{\beta}$,[OIII]\\ 1
6      &4:54:01.1 &-10:13:07.9 &73837        &89  \\
17      &4:53:59.7 &-10:12:19.9 &70699        &72  \\
18      &4:53:58.7 &-10:15:13.0 &74604        &63  \\
19      &4:53:57.9 &-10:12:11.1 &74846        &93  \\
20      &4:53:57.0 &-10:12:44.9  &73044       & 90 \\
21      &4:53:55.3 &-10:14:17.9  &75759        & 56 &[OII]\\
22      &4:54:06.2 &-10:15:53.1  &73317        &81 \\
23      &4:54:07.0 &-10:14:44.8 &71584        &79  \\
24      &4:54:07.9 &-10:15:36.4  &77596       & 38 \\
25      &4:54:08.8 &-10:14:33.9 &74727        &40  \\
26      &4:54:09.7 &-10:16:19.8 &75692        &94  \\
27      &4:54:10.5 &-10:14:52.6 &70522        &112 &[OII]\\
28      &4:54:13.1 &-10:15:09.7 &75002        &88 \\
29      &4:54:14.1 &-10:15:50.1 &73381        &57 \\
30      &4:54:14.7 &-10:15:37.7 &74874        &36 \\
31      &4:54:06.2 &-10:13:19.8 &74435        &42\\
32      &4:54:06.8 &-10:13:59.4 &73025        &51 \\
33      &4:54:07.7 &-10:13:44.9  &72150       & 61\\
34      &4:54:05.9 &-10:13:00.2 &73704        &93 \\
35      &4:54:09.3 &-10:14:10.7 &76736        &76 \\
36      &4:54:07.2 &-10:16:54.3  &73219       & 60  \\
37      &4:54:08.1 &-10:16:10.0  &70225       & 97 &[OII]\\
38      &4:54:15.4 &-10:16:52.5 &109645       &74  \\
39      &4:54:16.5 &-10:16:37.0 & 92925        &150 &[OII]\\
40      &4:54:18.5 &-10:16:51.4 & 88450        &85 &[OII]\\
41      &4:54:19.0 &-10:17:39.4 &75852        &52 \\
42      &4:54:15.4 &-10:16:14.4 &74055        &80 \\
43      &4:54:17.4 &-10:18:08.0 &75799        &62 \\
44      &4:54:19.1 &-10:16:47.3 &74073        &74\\
45      &4:54:15.8 &-10:14:07.2 &72967        &84\\
46      &4:54:19.9 &-10:13:22.7  &88820       & 41 &[OII]\\
47      &4:54:23.4 &-10:12:24.7 &75428        &86 \\
48      &4:54:24.3 &-10:12:01.1 &72048        &82 \\
49      &4:54:06.9 &-10:13:24.7 &74374        &47 \\
\hline
\end{tabular}
\end{flushleft}
\end{table*}
 
\begin{table*}
\caption[]{Mixture Model Parameters for Abell 521}
\begin{flushleft}
\begin{tabular}{llllllll}
\hline
{\bf group} &  {$N_g$} & {$\% N_{tot}$} &  {$\% L_{tot}
$}& { $x \pm \sigma_x$}& {$y \pm \sigma_y$}& {$m_{med}$}& {$m_{10-20}$}\\
&&&&  {(arcmin)}& {(arcmin)}&(V) &(V)\\
\hline
 {(1)}& {(2)}&  {(3)} &  {(4)} &  {(5)}
& {(6)} & {(7)} & {(8)}\\
\hline
1  & 184 & 46 & 46 &   1.8 $\pm$  1.7 &   0.9 $\pm$  2.4 & 21.0 & 19.3 \\
2  & 160 & 40 & 47 &  -1.6 $\pm$  1.4 &   0.0 $\pm$  2.3 & 20.9 & 19.4 \\
3  &  52 & 13 &  7 &  -2.4 $\pm$  1.0 &  -3.4 $\pm$  0.7 & 21.2 & 20.7 \\
\hline
\end{tabular}
\end{flushleft}
\end{table*}
 
\begin{table*}
\caption[]{ROSTAT analysis of velocity samples in Abell 521}
\begin{flushleft}
\begin{tabular}{llllll}
\hline
{\bf sample} & {\bf $N_{v}$} & {\bf $v_{BI}$}  & {\bf $S_{BI}$}  &
 AI  & TI\\
\hline
 All             &41  &$74132_{-249}^{+202}$ &$1386_{-139}^{+206}$ &-0.24 & 1.17
 \\
 KMM1            &9   &$74124_{-232}^{+289}$ &$806 _{-240}^{+682}$ &-0.65 & 1.74
 \\
 KMM2 (N)        &14  &$74122_{-650}^{+552}$ &$1994_{-226}^{+328}$ &-0.37 & 0.75
 \\
 KMM2 (S)        &5   &$73751_{-390}^{+228}$ &$747 _{-57} ^{+276}$ &-0.11 & ....\\
{\it red}        &19  &$74125_{-273}^{+218}$ &$1011_{-108}^{+214}$ &+0.49 & 1.19
 \\
{\it blue}       &20  &$73924_{-445}^{+408}$ &$1803_{-191}^{+256}$ &-0.09 & 0.96
 \\
\hline
\end{tabular}
\end{flushleft}
\end{table*}
 
\begin{table*}
\caption[]{Characteristics of the knots around the BCG in Abell 521}
\begin{flushleft}
\begin{tabular}{rrrrrrrrrr}
\hline
 knots   &   $R$  & $B$  & $V$  & $J$  & $B-R$ & $B-V$ & $V-J$ & v (km/s) &${\Delta}$~v~(\kms) \\
\hline
A     &   17.42  &   20.41  &  18.58  & 15.54  & 2.99  &  1.83  &  3.04 &74374&47\\
A1    &   19.27  &   22.24 &   20.40  & 16.88  & 2.97  &  1.84  &  3.52 &73857&55 \\
B     &   20.54 &    22.65  &  21.54  & 18.56  & 2.11  &  1.11  &  2.98 &75600&600\\
C     &   20.17  &   22.47  &  21.11  & 17.99  & 2.30  &  1.36  &  3.12 &74370&300\\
D     &   19.79  &   22.50  &  20.92  & 17.88  & 2.71  &  1.58  &  3.04 &74370&72\\
E      &  21.12  &   22.85  &  21.91  & 18.95  & 1.73  &  0.94  &  2.96 &

 &  \\
F      &  22.35   &  23.69  &  23.00  & 21.17  & 1.34   & 0.69  &  1.83 &
 &  \\
\hline
\end{tabular}
\end{flushleft}
\end{table*}
 

\begin{thebibliography}{99}
\bibitem{} Abell, G.O., 1958, ApJS, 3, 211
\bibitem{} Abell, G.O., Corwin, H.G., Olowin, R.P. 1989, ApJS. 70, 1 
\bibitem{} Arnaud, M., Maurogordato, S., Slezak, E., Rho, J., 1999, in preparation
\bibitem{} Beers T., Flynn K., Gebhart K., 1990, AJ 100, 32
\bibitem{} Beers T., Flynn K., Gebhart K., Huchra, J., Forman, W., Jones, C and
Bothun, G., 1992, ApJ, 400, 410
\bibitem{} Binggeli B., Tammann G.A., Sandage A., 1987, AJ 94, 251
\bibitem{} Bird, C.M., 1994, AJ 107, 1637
\bibitem{}Bird, C.M. and Beers, T., 1993, AJ, 107, 1637
\bibitem{}Bird, C.M., Davis, D., and Beers, T., 1995, AJ, 109, 920
\bibitem{} Bruzual, A.,G., Charlot, S., 1993, ApJ, 405, 538
\bibitem{}Burns, J., Roettiger, K., Pinkney, J.,Perley, R., Owen, F., and
Voges, W., 1995, Ap.J., 446, 583
\bibitem{} Cappi A., 1995, A\&A 301, 6
\bibitem{}Fort, B. and Mellier, Y., 1994, A\&ARv, 5, 239
\bibitem{}Frenk, C.,Evrard, A., White, S., Summers, F., 1996, 472, 460
\bibitem{}Gebhardt, K. and Beers, T.C.,  1991, ApJ, 383, 72
\bibitem{}Girardi, M., Fadda, D., Giuricin, G., Mardirossian, F., Mezzetti, M.,
Biviano, A.,  1996, ApJ, 457, 61
\bibitem{}Henry, J.P.,Briel, U.G., Nulsen, P.E.J., 1993, A\&A 271, 413
\bibitem{}Hoessel, J.G., Gunn, J.E., and Thuan, T.X.,  1980, ApJ, 241, 486
\bibitem{}Hoessel, J.G. and Schneider, D.P., 1985, AJ, 90, 1648
\bibitem{}Hausman, M.A., and Ostriker, J.P., 1978, ApJ, 224, 320
\bibitem{}Jones, B. J. T., \& Mazure, A. 1996, in Mapping  Measuring, and Modell
ing the
Universe, ASP Conference Series, 94, eds. P. Coles, V. J. Martinez \& M.-J.
Pons-Borderia (San Francisco:  Astronomical Society of the Pacific), p. 197
\bibitem{} Kowalski, M.P., Ulmer, M.P., Cruddace, 
R.G., Wood, K.S., 1984, ApJS 56, 403 
\bibitem{}Kriessler, J.R. and Beers, T.C.,  1997, AJ, 113, 80
\bibitem{} Le F\`evre O., Crampton C., Felenbok, P., and Monnet, G., 1994, $A\&A
$, 282, 325
\bibitem{} Le F\`evre O., Crampton C., Lilly S.J., Hammer F., Tresse L.,
 1995, ApJ, 455, 60
\bibitem{}Matthews, T.A., Morgan, W.W., and Schmidt, M., 1964, Ap.J.,140,35
\bibitem{}Mazure, A., Katgert, P., den Hartog, R., Biviano, A., Dubath, P., Esca
lera, E., Focardi, P., Gerbal, D., Giuricin, G., Jones, B., Le F\`evre, O., Mole
s, M., Perea, J., Rhee, G.,et al., 1996, $A\&A$, 310, 31
\bibitem{} Maurogordato S., Le Fevre, O., Proust, D., Vanderriest, C., Cappi,
 A., 1996, BCFHT, 34, 5
\bibitem{}Merritt, D. and Gebhardt,K., 1995, XXIXe Moriond Astrophysics Meetings
F.Durret, A.Mazure, J.Tran Thanh Van eds,
Editions Frontieres, p 11.
\bibitem{}Miller G.E., Scalo J.M., 1979, ApJS 41, 513
\bibitem{}Nottale L., 1983, A\&A 118, 85
\bibitem{}Oegerle, W., and Hill, J., 1994, AJ, 107, 857
\bibitem{} Pell\'o et al.,1999, in preparation
\bibitem{} Schindler, S., and Bohringer, H., 1993, $A\&A$, 269, 83
\bibitem{}Schombert, 1988, Ap.J., 328, 475
\bibitem{} Slezak et al.,1999, in preparation
\bibitem{} Tonry J., Davis, M., 1981, ApJ 84, 1511
\bibitem{}Tonry, J., 1985, AJ, 90, 2431
\bibitem{}Tully, R.B., Shaya, E.J., 1984, ApJ, 281, 31.
\bibitem{}Zabludoff, A., Huchra, J., Geller, M.J.,  1990, ApJS, 71, 1
\end{thebibliography}
\end{document}